\documentclass[12pt]{article}
\usepackage{bbm}
\usepackage{amsfonts}
\usepackage{amsmath}
\usepackage{latexsym}
\usepackage{natbib}
\usepackage{listings}
\usepackage{lscape}
\usepackage{graphicx}
\usepackage{setspace}
\usepackage{color}
\usepackage{multirow}
\usepackage{rotating}
\usepackage{epsfig}
\usepackage{booktabs}
\usepackage{tikz}
\usetikzlibrary{shapes,arrows}
\usepackage{topcapt}
\usepackage{lscape}

\usepackage{amsthm}

\newtheorem{definition}{Definition}
\newtheorem{theorem}{Theorem}
\newtheorem{lemma}{Lemma}
\newtheorem{proposition}{Proposition}

\newtheorem{remark}{Remark}
\newtheorem{condition}{Condition}

\title{Conditional Screening for Ultra-high Dimensional Covariates with Survival Outcomes}
\author{Hyokyoung~Grace~Hong\thanks{Hyokyoung~Grace~Hong is Assistant Professor in the Department of Probability and Statistics, Michigan State University, East Lansing, MI 48823.  Jian~Kang is Assistant Professor in the Department of Biostatistics, University of Michigan, Ann Arbor, MI 48019. Yi Li is Professor of Biostatistics, Director of Kidney Epidemiology and Cost Center, University of Michigan, Ann Arbor, MI 48019. },~Jian~Kang\thanks{To whom correspondence should be addressed: jiankang@umich.edu. 
This research was partially supported by a grant from NSA (H98230-15-1-0260, Hong), an NIH  grant (R01MH105561, Kang) 
and  Chinese Natural Science Foundation (11528102, Li).}~~and~Yi~Li}
\date{}                                           

\newcommand{\cm}[1]{\ignorespaces}

\newcommand\jian[1]{{\color{black}{#1}}}
\newcommand\grace[1]{{\color{black}{#1}}}

\def\bfa{\mathbf a}

\def\bfv{\mathbf v}

\def\bfxi{\boldsymbol \xi}

\def\bfA{\mathbf A}
\def\bfB{\mathbf B}

\def\bfI{\mathbf I}

\def\bfZ{\mathbf Z}

\def\bfV{\mathbf V}
\def\bfU{\mathbf U}
\def\bfW{\mathbf W}

\def\bfalpha{\boldsymbol \alpha}
\def\bfbeta{\boldsymbol \beta}

\def\bfzeta{\boldsymbol\zeta}
\def\bfzero{\boldsymbol 0}
\def\bfone{\boldsymbol 1}

\def\cB{\mathcal B}
\def\cC{\mathcal C}

\def\cM{\mathcal M}

\def\cF{\mathcal F}
\def\mCov{\mathrm{Cov}}
\def\mVar{\mathrm{Var}}

\def\mE{\mathrm{E}}

\def\mP{\mathrm{P}}

\def\mbR{\mathbb R}

\def\md{\mathrm d}

\def\mE{\mathrm{E}}

\DeclareMathOperator*{\argmin}{\mathrm{argmin}}

\def\rT{\mathrm T}

\begin{document}

\bibliographystyle{asa}
\bibpunct{(}{)}{,}{a}{}{;}

\maketitle
\begin{abstract}
 Identifying important biomarkers that are predictive for cancer patients' prognosis is key in  gaining  better insights into the biological influences on the disease and 
 has become a critical component  of  precision medicine.
The emergence of large-scale biomedical survival studies, which typically involve excessive number of biomarkers, has brought high demand in designing efficient  screening tools for selecting predictive biomarkers.  The vast amount of biomarkers
defies any existing variable selection methods via regularization. The recently developed variable screening methods, though powerful in many practical setting,  fail to incorporate  prior information on the importance of each biomarker and are less powerful in detecting marginally weak while jointly 
important signals.  We propose a new conditional screening method  for survival outcome data by computing the marginal contribution of each biomarker given priorly known biological information. This is based on the premise that some biomarkers are known to be associated with 
disease outcomes a priori. Our method possesses  sure screening properties and a vanishing false selection rate. The utility of the proposal is further confirmed with extensive simulation studies and analysis of a Diffuse large B-cell lymphoma (DLBCL) dataset. 

\medskip


\end{abstract}

\newpage
\setlength{\baselineskip}{24pt}		
\section{Introduction}
\label{intro}


Despite much progress made in the past two decades,  many cancers do not have a proven means of prevention or effective treatments. 
Precision medicine that takes into account individual susceptibility has become a valid approach to gaining better insights into the biological influences on cancers, which is expected to benefit  millions of cancer patients. A critical component of precision medicine lies in  detecting and  identifying important biomarkers that are predictive for cancer patients' prognosis. The emergence of large-scale biomedical survival studies, which typically involve excessive number of biomarkers, has brought high demand in designing efficient  screening tools for selecting predictive biomarkers.   \grace{The presented work is motivated by a genomic study of Diffuse large B-cell lymphoma (DLBCL) \cite{Rosenwald:2002},
 with the goal of  identifying gene signatures out of  7399 genes for  predicting  survival
among 240 DLBCL patients. The results may address whether the DLBCL patients' survival after chemotherapy could be regulated by the molecular features.}

 The recently developed variable screening methods, such as the sure independence screening  proposed by~\cite{FanLv2008},  have emerged as a powerful tool to solve this problem,  but their validity often hinges upon the partial faithfulness assumption, that is,
 the jointly important variables are also marginally important. Consequently, they will fail to identify the  \emph{hidden} variables that are jointly important but have weak marginal associations with the outcome,  resulting  in poor understanding of the molecular mechanism underlying or regulating the disease.  To alleviate this problem, ~\cite{FanLv2008} 
further suggested an  iterative  procedure {(ISIS)} by repeatedly using the residuals from the previous iterations,
which has gained much popularity. However, 
the required iterations have  increased the computational burden, and the statistical properties  are elusive.


On the other hand,  intensive biomedical research has generated  a large body of biological knowledge. For example,  \grace{several  studies have confirmed AA805575,  a
Germinal-center B-cell signature gene,  is relevant to DLBCL survival \citep{Gui:2005, Liu:2013}.}  Including such prior knowledge for improved accuracy in  variable selection has drawn much
interest.   \cite{Barut:2015}
proposed a  conditional screening (CS) approach in the framework of a generalized linear model (GLM) when some prior knowledge on feature selection is known, and  showed that the CS
approach provides a powerful means to identify jointly-informative but  marginally weak associations, 
and \cite{Hong:2015} further proposed to integrate prior information using data-driven approaches. 


 Development of high dimensional screening tools with survival outcome  has  been fruitful.  Some related work includes  an (iterative) sure screening procedure for  Cox's proportional hazards model~\citep{Fan:2010},  a marginal maximum partial likelihood estimator (MPLE) based screening procedure~\citep{DZhao:2012}, a censored rank independence screening method which is robust to outliers and applicable to a general class of survival model~\citep{Song:2014}.  But to the best of our knowledge, all these methods essentially posit the partial faithfulness assumption and do not incorporate  the known prior biological information. As a result, they will be likely to suffer the inability  to identify marginally weak but jointly important signals.


To fill the gap,  we propose a new conditional screening method for the Cox proportional hazards model by computing the marginal contribution of each covariate given priorly known information. \jian{We refer to it as Cox conditional screening (CoxCS).} As opposed to the conventional  marginal screening methods, our method enables the detection of marginally weak  but jointly important signals, which will have important biological applications as shown in the data example section.  Moreover, in contrast with most screening methods that usually employ subjective thresholds for screening,  \jian{we also propose a principled cut-off to govern the screening and  control the false positives in light of~\cite{DZhao:2012}. This will be especially important in the presence of hidden variables.}

\begin{figure}[htbp] 
   \centering
   \includegraphics[width=6.0in]{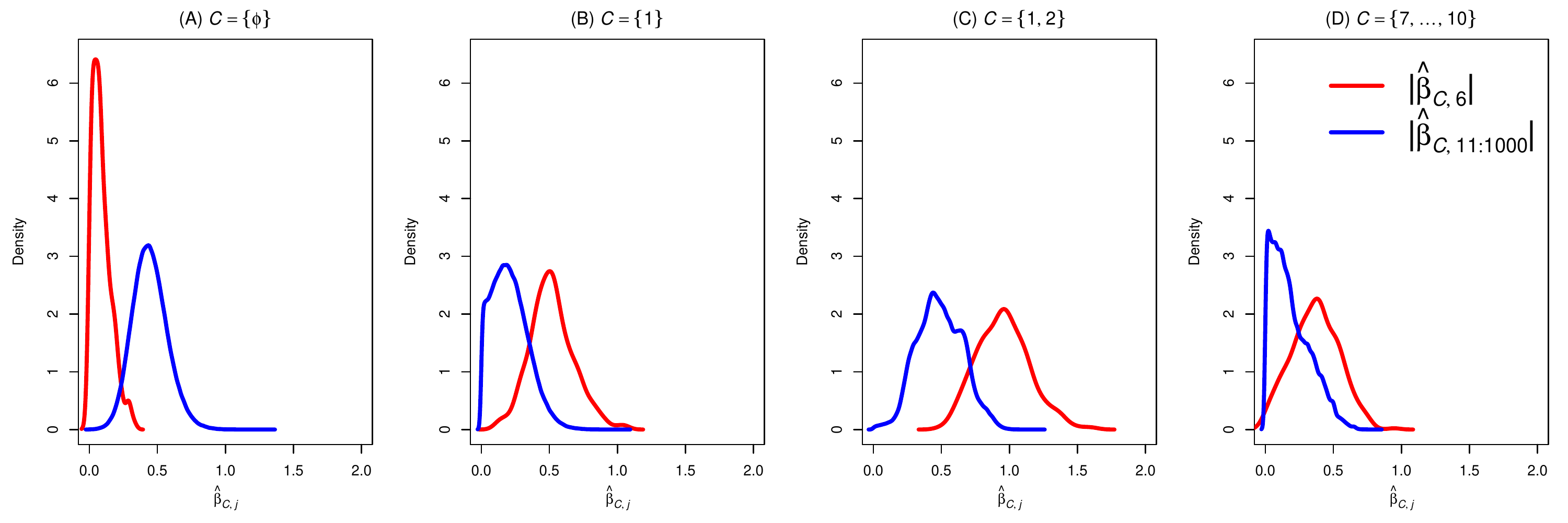} 
   \caption{\jian{Density of the screening statistics $|\widehat \beta_{\cC,6}|$ (red) for the hidden variable compared with a mixture of densities of screening statistics $|\widehat \beta_{\cC,11:1000}|$ (blue) for the noise variables with different conditioning sets: (A) $\cC=\{\emptyset\}$ which is equivalent to marginal screening; (B) $\cC=\{1\}$ one truly active variables; (C) $\cC=\{1,2\}$ two truly active variables; (D) $\cC=\{7,\ldots, 10\}$ four noisy variables.}}
   \label{fig:den}
\end{figure}

\jian{To demonstrate the utility of  CoxCS  in recovering important hidden variables, we consider an example with 100 subjects and 1,000 covariates, where the survival times were generated from a Cox model with baseline hazards function being $1$, and the covariates being generated from the multivariate standard normal distribution with equal correlation 0.5. The true coefficients in the Cox model are set to be $\beta_1 = \ldots = \beta_5 = 1$ and $\beta_6 = -2.5$ and $\beta_j = 0$ for $j \in\{7,\ldots, 1000\}$. By design, variable 6 is the hidden variable in that it is marginally uncorrelated with survival times approximately; see Example 1 in Section 4 for more details.  Let $\widehat \beta_{\cC,j}$ denote the screening statistics by the CoxCS approach (defined in Section 2), where $\cC$ indexes variables that are pre-included into the model. When $\cC = \emptyset$, the CoxCS is equivalent to the marginal screening approach for the Cox proportional hazard model. Figure \ref{fig:den} summarizes the densities of the screening statistics for the hidden variable $6$  and noisy variables 10 to 1,000 for different sets of conditional variables based on 400 simulated datasets.  The results show that  with a high probability the marginal screening statistic for hidden variable 6 is much smaller than those of noisy variables. When the conditioning set includes one truly active variable, the density plots show a clear separation between the hidden variables and the noisy variables. When we include more truly active variables, this separation becomes larger. Interestingly, when conditional on noisy variables that are correlated with both active and hidden variables
  in the model, the chance of identifying the hidden variable using CoxCS is still higher than the marginal screening. A similar phenomenon was  observed in the GLM setting \citep{Barut:2015}. \grace {This is because  when  such ``noisy'' variables are correlated with both marginally important  variables and hidden variables, they may effectively function as surrogates for the active variables and conditioning on them can help detect hidden variables.}}

\jian{The theory of conditional screening for GLM has been established by~\cite{Barut:2015}. But its extension to the survival context is challenging and elusive, calling for new techniques. To this end, we propose two new functional operators on random variables to characterize their linear associations given other random variables: the conditional linear expectation and the conditional linear covariance. Both are critical to formulate the regularity conditions for the population level properties of CoxCS with statistically meaningful interpretations, and facilitate the development of theory for conditional screening approaches in general settings. A similar concept of the conditional linear covariance has been introduced by~\cite{Barut:2015}, but it can not be used for the survival outcome data.} 
\jian{In summary,  the proposed method is  computationally efficient, adapts  to sparse and weak signals, enjoys the good theoretical properties under weak regularity conditions, and works robustly in a variety of settings to identify hidden variables. }

\jian{The remaining of this paper is organized as follows. In Section 2, we review the Cox proportional hazard model and present CoxCS approach with some alternatives. In Section 3, we list the regularity conditions and establish the sure screening properties. In Section 4, we further  conduct simulation studies to compare our method with the major competing methods under under a number of scenarios. In  Section 4, we apply our method to study  the DLBCL data.  We conclude with a a brief discussion on the future work in Section 5. }

\section{Model}
\label{model}

\subsection{The Cox Proportional Hazard Model}
Suppose we have $n$ observations with $p$ covariates. Let $i$ and $j$ respectively index subjects and covariates. Denote by $Z_{i,j}$ covariate $j$ for subject $i$, write $\bfZ_i = (Z_{i,1},\ldots, Z_{i,p})^{\rT}$. Let $T_i$ be the underlying survival time and $C_i$ be the censoring time. We observe $X_i = \min\{T_i, C_i\}$, and $\delta_i = I[T_i \leq C_i]$, where $I(\cdot)$ is the indicator function.  Assume that there exists $\tau>0$, such that $\mP(X_i>\tau \mid \bfZ_i) = 0$ and assume that the event time $T_i$ and the censoring time $C_i$ are independent. Suppose $T_i$ follows a Cox proportional hazards model 
\begin{eqnarray}\label{eq:Cox_model}
\lambda(t; \bfZ_i) = \lambda_0(t) \exp(\bfalpha^{\rT} \bfZ_i),\label{eq:hazard}
\end{eqnarray}
where $\lambda_0(t)$ is an-unspecified baseline hazard and $\bfalpha = (\alpha_{1},\ldots, \alpha_{p})^{\rT}$ is the true-coefficient. Let $\Lambda_0(t) = \int_0^t \lambda_0(s)\md s$ be the cumulative hazard function.   Suppose there is a set of covariates that are known a priori to be related to the survival outcome. Denote by $\cC$ the indices of these covariates. Let $q = |\cC|$ be the number of covariates in $\cC$.   Write $\bfZ_{i,\cC} = (Z_{i,j},j\in\cC)^{\rT}$, $\bfZ_{i,-\cC} = (Z_{i,j},j\notin\cC)^{\rT}$,  $\bfalpha_{\cC} = (\alpha_{j}, j\in \cC)^{\rT}$ and $\bfalpha_{-\cC} = (\alpha_{j}, j\notin \cC)^{\rT}$. Note that  in our problem, $\cC$ is known but both $\bfalpha_{\cC}$ and $\bfalpha_{-\cC}$ are unknown. Then the true hazard function in \eqref{eq:hazard} is equivalent to 
\begin{eqnarray}
\lambda(t; \bfZ_i) = \lambda_0(t) \exp(\bfalpha_{\cC}^{\rT} \bfZ_{i,\cC} +\bfalpha_{-\cC}^{\rT} \bfZ_{i,-\cC}),\label{eq:hazard_C}
\end{eqnarray}

To estimate $\bfalpha_{\cC}$ and $\bfalpha_{-\cC}$, we introduce the independent counting process $N_i(t) = I(X_i \leq t, \delta_i = 1)$ and the at-risk process $Y_i(t) = I[X_i \geq t]$. When $p$ is small, we can obtain the partial likelihood estimator $\widehat\bfalpha = (\widehat\bfalpha^{\rT}_{\cC}, \widehat\bfalpha^{\rT}_{-\cC})^{\rT} $ by solving the estimation equation $\bfU(\bfalpha) = \bfzero_p$ with $\bfU(\bfalpha) = (U_{1}(\bfalpha),\ldots, U_{p}(\bfalpha))^{\rT}$ and 
\begin{eqnarray}\label{eq:est_eqn}
U_{j}(\bfalpha) = \sum_{i=1}^n \int_{0}^\tau\left\{Z_{i,j} - \frac{S_{j}^{(1)}(t,\bfalpha)}{S_{j}^{(0)}(t, \bfalpha)}\md N_i(t)\right\},
\end{eqnarray}
with
\begin{eqnarray}\label{eq:def_S}
S^{(m)}_{j}(t,\bfalpha) = \frac{1}{n}\sum_{i=1}^n Z^m_{i,j} Y_i(t)\exp(\bfalpha_{\cC}^{\rT} \bfZ_{i,\cC}+\bfalpha_{-\cC}^{\rT} \bfZ_{i,-\cC}),
\end{eqnarray}
for $m \in \{0,1,2,\ldots,\}$.
When $p>n$, it is computationally and theoretically infeasible to directly solve the equation \eqref{eq:est_eqn}. By imposing sparsity on the coefficients,  one may maximize the penalized partial likelihood to obtain solutions. However, when $p>>n$, we need to employ a variable screening procedure first before performing
regularized regression. We propose a new conditional screening procedure in the next section. 

\subsection{Cox Conditional Screening}
We fit the marginal Cox regression by including the known covariates in $\cC$. Specifically, for $j \notin \cC$, we have the following marginal Cox regression model
\begin{eqnarray}\label{eq:marginal}
\lambda_{j}(t; \bfZ_i) = \lambda_{j,0}(t) \exp(\bfbeta_{\cC}^{\rT} \bfZ_{i,\cC} +\beta Z_{i,j}),
\end{eqnarray}
from which the maximum partial likelihood estimation equation $(\widehat\bfbeta_{\cC,j}^{\rT},\widehat\beta_j)^{\rT}$ can be obtained. It is given by the solution of the following equation:
\begin{eqnarray}\label{eq:est_eqn}
\bfV_j(\bfbeta_{\cC},\beta) =  [V_{j,k}(\bfbeta_{\cC},\beta), k \in\cC\cup \{j\}]^{\rT} = \bfzero_{q+1},
\end{eqnarray}
with 
\begin{eqnarray}\label{eq:element_eqn}
{ V_{j,k}(\bfbeta_\cC,\beta) = \sum_{i=1}^n \int_0^\tau\left\{Z_{i,k} - \frac{R_{j,k}^{(1)}(\bfbeta_{\cC},\beta, t)}{R_{j,k}^{(0)}(\bfbeta_{\cC},\beta, t)}\md N_i(t)\right\},}
\end{eqnarray}
and
\begin{eqnarray}\label{eq:def_R}
{R^{(m)}_{j,k}(\bfbeta_{\cC},\beta,t) = \frac{1}{n}\sum_{i=1}^n Z^{m}_{i,k} Y_i(t) \exp(\bfbeta_{\cC}^{\rT} \bfZ_{i,\cC}+\beta Z_{i,j})},
\end{eqnarray}
for $k \in \cC\cup \{j\}$ and $m \in \{0, 1, 2, \ldots \}$. For a given threshold $\gamma>0$. The selected index set in addition to set $\cC$ is given by
\begin{eqnarray}\label{eq:select_idx}
\widehat\cM_{-\cC} = \left\{j\notin \cC: |\widehat\beta_j|\geq \gamma\right\}. 
\end{eqnarray}
Namely, we recruit variables with large additional contribution given $\bfZ_{\cC}$. \grace{We refer to this method as  Cox conditional screening (CoxCS).}

\section{Theoretical Results}\label{theoretic}
We establish the theoretical properties of the proposed methods by introducing a few new definitions along with the basic properties. 
\subsection{Definitions and Basic Properties}
Let $(\Omega, \cF, \mP)$ be the probability space for all random variables  introduced in this paper, where $\Omega$ is the sample space, $\cF$ is the $\sigma$-algebra as the set of events and $\mP$ is the probability measure. Let $\mbR^d$   be a $d$-dimensional Euclidean vector space, for positive integer $d$. Denote by $\mE[\bullet]$, $\mVar[\bullet]$ and $\mCov[\bullet,\bullet]$ the commonly used expectation, variance and covariance operator in the  probability theory, respectively.  For any $d\geq 1$, any random variable $\bfxi: \Omega\to \mbR^d$ and any operator $A$, denote  by $A[\bullet \mid \bfxi]$ conditional $A$ of $\bullet$ given $\bfxi$.  For any vector $\bfa = (a_1,\ldots, a_p) \in \mbR^p$, let $\bfa_{\cC} = (a_j, j\in\cC)^{\rT}$ be the sub vector where all its elements are indexed in $\cC$. Let $\|\bfa\|_d = \sqrt[d]{\sum_{j=1}^p |a|^d_j}$ be the L-$d$ norm for any vector $\bfa\in\mbR^p$. For a sequence of random variables indexed by $\{\xi_n\}$, $\xi_n = o_p(1)$ if and only if for any $\epsilon_1>0$ and $\epsilon_2>0$, there exists $N$ such that for any $n>N$ $\mP[|\xi_n|>\epsilon_1]<\epsilon_2$. 

For simplicity, let $T$, $C$, $X$, $Y(t)$, $Z_{j}$, $\bfZ$ and $\delta$ represent $T_i$, $C_i$, $X_i$, $Y_i(t)$, $Z_{i, j}$, $\bfZ_i$ and $\delta_i$ respectively, by removing the subject index $i$.  Let $S_T(t\mid \bfZ)$ and $S_C(t)$ represent the survival functions for the event time $T$ and censored time $C$. Let $F_T(t\mid \bfZ) = 1 - S_T(t\mid \bfZ)$. 

\begin{definition}\label{def:true_set}
Let $\cM_{-\cC} = \{j\notin\cC, \alpha_{j} \neq 0\}$ and $w = \sum_{j\notin \cC} I[\alpha_j \neq 0]$ be the true set of non-zero coefficients and its cardinality in model \eqref{eq:Cox_model}, aside from the important predictors known a priori. 
\end{definition}

To study the asymptotic property of $(\widehat\bfbeta_{\cC,j}^{\rT},\widehat\beta_j)^{\rT}$, define the population level quantity as follows:

\begin{definition}\label{def:pop_solution}
Let $(\bfbeta_{\cC,j}^{\rT},\beta_{j})^{\rT}$ be the solution of the following equations
\begin{eqnarray}\label{eq:def_v}
\bfv_j(\bfbeta_{\cC},\beta) = [v_{j,k}(\bfbeta_{\cC},\beta), k\in \cC\cup\{j\} ]^{\rT} = \bfzero_{q+1},
\end{eqnarray}
with
\begin{eqnarray}\label{eq:def_v}
v_{j,k}(\bfbeta_{\cC},\beta) = \int_0^\tau \left[s^{(1)}_k(t) -\frac{r^{(1)}_{j,k}(t,\bfbeta_{\cC},\beta)}{r^{(0)}_{j,k}(t, \bfbeta_{\cC},\beta)} s^{(0)}_k(t) \right] \md t, 
\end{eqnarray}
where $s^{(m)}_{k}(t) = \mE\left[Z_{k}^m\md N(t)\right] $ and $r^{(m)}_{j,k}(t,\bfbeta_{\cC},\beta) = \mE[R^{(m)}_{j,k}(t,\bfbeta_{\cC},\beta)] $.
\end{definition}

\begin{definition}\label{def:cond_solution}
Let $\bfbeta_{\cC,0}$ be the solution of the following equations
\begin{eqnarray}\label{eq:cond_solution}
\bfv_{\cC}(\bfbeta_{\cC}) = [v_{j,k}(\bfbeta_{\cC},0), k\in \cC]^{\rT} = \bfzero_{q}.
\end{eqnarray}
\end{definition}

\begin{proposition}\label{prop:solution}
$\bfv_j(\bfbeta_{\cC,0},0)=\bfzero_{q+1}$ if and only if $v_{j,j}(\bfbeta_{\cC,0},0) = 0$, for all $j\in\cC$.
\end{proposition}

To understand the intuition of  the population level properties for the Cox conditional screening, we need to define a conditional linear expectation.   A similar concept has been used to study the conditional sure independence screening (CSIS) in the GLM setting by~\cite{Barut:2015}. We provide a formal definition here.
\begin{definition}\label{def:cond_linear_exp}
For two random variables $\bfzeta: \Omega\to \mbR^d$ and  $\bfxi: \Omega\to \mbR^p$. The conditional linear expectation of $\bfzeta$ given $\bfxi$ is defined as
\begin{eqnarray}\label{eq:cond_linear_exp}
\mE^*(\bfzeta \mid \bfxi) =\mE[\bfzeta] + \bfA^{\rT}\{\bfxi-\mE(\bfxi)\},
 \end{eqnarray}
 where $\bfA = \argmin_{\bfB \in \mbR^d \times \mbR^p} \mE[(\bfzeta - \mE[\bfzeta] - \bfB^{\rT}\{\bfxi-\mE(\bfxi)\})^2 \mid \bfxi]$. Also, define notation $\mE^*(\bfzeta) = \mE(\bfzeta)$. 
\end{definition}
The basic properties of the conditional linear expectation are listed in the following proposition. 

\begin{proposition}\label{prop:CLE}
Let $\bfzeta$, $\bfzeta_1$, $\bfzeta_2$ and $\bfxi$ be any four random variables in the probability space $(\Omega,\cF, \mP)$. The following properties hold for the conditional linear expectation $\mE^*[\bullet \mid \bfxi]$ given $\bfxi$: 
\begin{enumerate}
\item Closed form: $\mE^*(\bfzeta \mid \bfxi) =\mE[\bfzeta] + \mCov(\bfzeta,\bfxi)\mVar[\bfxi]^{-1} \{\bfxi-\mE(\bfxi)\}$. 
\item Stability: $\mE^*[\bfxi \mid \bfxi] = \bfxi$. 
\item Linearity:  $\mE^*[\bfA_1\bfzeta_1+\bfA_2\bfzeta_2 \mid \bfxi] = \bfA_1\mE^*[\bfzeta_1 \mid \bfxi]+\bfA_2\mE^*[\bfzeta_2 \mid \bfxi]$, where $\bfA_1$ and $\bfA_2$ are two matrices that are compatible  with the equation. 
\item Law of total expectation: $\mE^*[\mE^*(\bfzeta \mid \bfxi)] = \mE[\mE^*(\bfzeta \mid \bfxi)] = \mE[\bfzeta]$.
\end{enumerate}
\end{proposition}
\begin{remark}
In general, $\mE^*(\bfzeta \mid \bfxi) \neq \mE(\bfzeta \mid \bfxi)$. Also, $\bfzeta$ and $\bfxi$ are independent does not imply $\mE^*(\bfzeta \mid \bfxi) = 0$, unless $\bfzeta$ and $\bfxi$ are jointly normally distributed. 
\end{remark}
\begin{definition}\label{def:cond_linear_cov}
For any random variables $\bfzeta_1: \Omega \to \mbR^{d_1}$, $\bfzeta_2: \Omega \to \mbR^{d_2}$ and $\bfxi: \Omega \to \mbR^{p}$. The conditional linear covariance between $\bfzeta_1$ and $\bfzeta_2$ given $\bfxi$ is defined as
\begin{eqnarray}\label{eq:cond_linear_cov}
\mCov^*(\zeta_1,\zeta_2 \mid \bfxi) =  \mE^*[\{\zeta_1 - \mE^*(\zeta_1 \mid \bfxi)\}\{\zeta_2 - \mE^*(\zeta_2 \mid \bfxi)\} \mid \bfxi].
\end{eqnarray}
\end{definition}
\begin{remark}
By Proposition \ref{prop:CLE}, we can easily verify the following properties.
\begin{proposition}\label{prop:cond_linear_cov}
The conditional linear covariance defined in definition~\ref{def:cond_linear_cov} has the following properties
\begin{enumerate}
\item Linear independence and linear zero correlation:
$$\mCov^*(\bfzeta_1,\bfzeta_2 \mid \bfxi) = 0\qquad \Leftrightarrow \qquad \mE^*(\bfzeta_1 \bfzeta_2 \mid \bfxi) = \mE^*(\bfzeta_1 \mid \bfxi) \mE^*(\bfzeta_2 \mid \bfxi).$$ 
\item Expectation of conditional linear covariance:
$$\mE[\mCov^*(\bfzeta_1,\bfzeta_2\mid \bfxi)] = \mCov(\bfzeta_1,\bfzeta_2) -  \mCov(\bfzeta_1,\bfxi) \mVar(\bfxi)^{-1} \mCov(\bfxi,\bfzeta_2).$$ 
\item Sign: for any increasing function $h(\cdot): \mbR \to \mbR$ and  random variable $\eta: \Omega \to \mbR$, then 
$$\mCov^*(h(\eta),\eta\mid \bfxi) \geq 0. $$
\end{enumerate}
\end{proposition}
\end{remark}

\begin{definition}\label{def:cond_exp_v}
Define 
$$v_j(\bfbeta_{\cC},\beta) = v_{j,j}(\bfbeta_{\cC},\beta) +\sum_{k\in\cC} a_k v_{j,k}(\bfbeta_{\cC},\beta),$$ where vector $\bfa_\cC  = [a_k, k\in \cC]^{\rT}$ such that $\mE^*[Z_j \mid \bfZ_{\cC}] = \sum_{k\in \cC} a_k Z_{k}$. 
\end{definition}

\begin{proposition}\label{prop:linear_comb_solution}
$v_{j,j}(\bfbeta_{\cC,0},0) = 0$ if and only if $v_j(\bfbeta_{\cC,0},0)  = 0$. 
 \end{proposition}

\subsection{Regularity Conditions}
We list all conditions for the theoretical results. 
\begin{condition}\label{cond:R}
For each $j\notin \cC$ and $k\in\cC\cup\{j\}$, there exists a neighborhood of $(\bfbeta_{\cC,j}^{\rT}, \beta_j)^{\rT}$, which is defined as 
$$\cB_j = \{(\bfbeta_{\cC}^{\rT}, \beta)^{\rT}: \|(\bfbeta_{\cC}^{\rT}, \beta_j)^{\rT}-(\bfbeta_{\cC,j}^{\rT}, \beta_j)^{\rT}\|_1<\delta_j\},\qquad\mbox{with}\qquad \delta_j>0,$$ such that for each $\tau<\infty$,
\begin{enumerate}
\item For $m = 0, 1$, 
$$\sup_{t\in[0,\tau], (\bfbeta_{\cC}^{\rT}, \beta)^{\rT}\in \cB_j}\|R_{j,k}^{(m)}(\bfbeta_{\cC},\beta) - r_{j,k}^{(m)}(\bfbeta_{\cC},\beta)\|_2 \to 0,$$
in probability as $n\to \infty$. 
\item There exists a constant $L>0$ such that 
$$L = \min_{j\notin\cC}\left[\inf_{t\in [0,\tau], (\bfbeta_{\cC}^{\rT},\beta)^{\rT}\in \cB_j}\{r_{j,k}^{(0)}(\bfbeta_{\cC},\beta,t)\}\right].$$
\end{enumerate}
Of note, $\{r_{j,k}^{(0)}(\bfbeta_{\cC},\beta,t)\}$ does not depend on $k$. Let $\delta = \max_{j\notin\cC} \delta_j$. 
\end{condition}

\begin{condition}\label{cond:covariate}
The covariates $Z_j$'s satisfy the following conditions
\begin{enumerate}
\item For $j \in \{1,\ldots, p\}$, $\mE[Z_j] = 0$ and there exists a constant $K_0$ such that $\mP(Z_j > K_0) = 0$. 
\item $Z_j$ is a time constant variable, for all $j$. 
\item All $Z_j$'s, $j \in \cM_{-\cC}$ are independent of all $Z_{k}$'s,  $k \notin \cM_{-\cC}$ given $\bfZ_{\cC}$. 
\item For constant $c_1>0$ and $\kappa < 1/2$, 
$$\min_{j\in\cM_{-\cC}}\left| \mE[\mCov^*(Z_j, \mP[\delta = 1\mid \bfZ]\mid \bfZ_{\cC})])\right| \geq c_1 n^{-\kappa}.$$
\end{enumerate}
\end{condition}

\begin{condition}\label{cond:beta_alpha}
There exists a constant $K_1$ such that
$$\|\bfalpha\|_1 < K_1\ \mbox{and}\ \|(\bfbeta^{\rT}_{\cC,j},\beta_j)^{\rT}\|_1 < K_1,$$
for all $p>0$.
\end{condition}

\begin{condition}\label{cond:beta_hat}
For all $j\notin\cC$, there exists a constant $M>0$ such that 
$$M\|(\widehat\bfbeta_{\cC,j}^{\rT},\widehat\beta_j)^{\rT} -\bfbeta_{\cC,j}^{\rT},\beta_j)^{\rT}  \|_{2}\leq \|\overline\bfV_j(\widehat\bfbeta_{\cC,j},\widehat\beta_j)-\overline\bfV_j(\bfbeta_{\cC,j},\beta_j)\|_2.$$
\end{condition}

\subsection{Properties on Population Level}
\begin{lemma}\label{lem:unique}
The solution of $\bfv_j(\bfbeta_{\cC},\beta) = \bfzero_{q+1}$ and the solution of $\bfv_{\cC}(\bfbeta_{\cC}) = \bfzero_q$ are both unique, for any $j \notin \cC$. 
\end{lemma}

\begin{theorem}\label{thm:ident}
Suppose Condition \ref{cond:covariate} hold, $\beta_{j} = 0$ if and only if $\alpha_{j} = 0$ for all $j\notin\cC$. 
\end{theorem}

\begin{theorem}\label{thm:beta_low_bound}
Suppose Condition \ref{cond:covariate}  holds. There exist constants $c_2>0$ and $\kappa<1/2$ such that 
$$\min_{j\in\cM_{-\cC}}|\beta_{j}| \geq c_2 n^{-\kappa}.$$
\end{theorem}

\subsection{Properties on Sample Level}
\begin{theorem}\label{thm:prob_bound}
Suppose Conditions~\ref{cond:R}--\ref{cond:beta_hat} hold. For any $\epsilon_1 > 0$ and any $\epsilon_2>0$, there exits positive constants $c_3$, $c_4$ and integer $N$ such that for any $n>N$,  
\begin{enumerate}
\item For any $0<\kappa<1/2$, 
\begin{eqnarray*}\label{eq:prob_bound}
\mP\left[\max_{j \in\cM_{-\cC}}| \widehat\beta_{j} - \beta_j| >\frac{c_2}{2}(n^{-\kappa} - \epsilon_1)\right] \leq 2w(q+1)\exp(-c_3 n^{1-2\kappa})+\epsilon_2.
\end{eqnarray*}
where $w$ is the size of $\cM_{-\cC}$, $q$ is the size of $\cC$ and $c_2$ is the same value in Theorem~\ref{thm:beta_low_bound} and $c_3$ does not dependent on $\epsilon_1$, $\epsilon_2$ and $\kappa$, but $N$ depends on $\epsilon_1$ and $\epsilon_2$. 
\item If $\gamma_n = c_4 n^{-\kappa}$, where $\kappa$ is the same number in Condition \ref{cond:covariate}, then
\begin{eqnarray}
\mP\left[\min_{j\in \cM_{-\cC}}|\widehat \beta_j| > \gamma_n \right] \geq 1 -2w(q+1)\exp(-c_3 n^{1-2\kappa}) - \epsilon_2. 
\end{eqnarray}
where $c_4$ does not depend on $\epsilon_1$, $\epsilon_2$ and $\kappa$, thus
\begin{eqnarray}\label{eq:prob_bound}
\lim_{n\to\infty}\mP\left[\cM_{-\cC} \subseteq \widehat\cM_{-\cC}\right]  = 1. 
\end{eqnarray}
\end{enumerate}

\end{theorem}



%
%

\subsection{Controlling the False Discover Rate}
Define the information matrix
\begin{eqnarray}\label{eq:info_mat}
\bfI_j(\beta_{\cC},\beta_j) = - \left(\frac{\partial V_{j,k}(\bfbeta_{\cC},\beta_j)}{\partial \beta_{k'}}\right)_{k,k'\in\cC\cup \{j\}},
\end{eqnarray}
which is of $q+1$ dimension.
Denote $\widehat\sigma^2_j = [\bfI_j(\beta_{\cC,j},\widehat\beta_j)]^{-1}_{q+1,q+1}$ be the variance estimate of $\widehat\beta_j$. For a given threshold $\gamma>0$, we can have a different way to select the index which is given by
\begin{eqnarray}\label{eq:select_idx1} 
\widehat\cM^*_{-\cC} = \left\{j\notin \cC: \frac{|\widehat\beta_j|}{\widehat\sigma_j}\geq \gamma\right\},
\end{eqnarray}
 as suggested by~\cite{DZhao:2012}. \grace{We refer to~\eqref{eq:select_idx1} as ``CS-Wald".}

\grace{Another alternative to construct the screening statistics, which is also scale free, is to utilize the partial log likelihood ratio statistic.
Specifically,
\begin{eqnarray}
\label{LRT}
\ell(\bfbeta_\cC,\beta) = \sum_{i=1}^n \int_0^\tau\left\{ \bfbeta_{\cC}^{\rT} \bfZ_{i,\cC}+\beta Z_{i,j} - \log \left(R_{j,k}^{(0)}(\bfbeta_{\cC},\beta,t)\right)\md N_i(t)\right\}.
\end{eqnarray}
}

\grace {Suppose   $(\widehat\bfbeta_{\cC,j}^{\rT},\widehat\beta_j)^{\rT}$  maximizes (\ref{LRT}) for a given $j$. 
Then, for a given threshold $\gamma>0$, the index set can be chosen by considering the following likelihood ratio statistic.
\begin{eqnarray}\label{eq:select_idx2}
\widehat\cM^*_{-\cC} = \left\{j\notin \cC: \ell (\hat{\bfbeta}_{\cC,j},\hat{\beta_j})-\ell (\hat{\bfbeta}_{\cC,0}, \beta=0)\geq \gamma \right\},
\end{eqnarray}
where $\hat{\bfbeta}_{\cC,0}$ maximizes $\ell(\bfbeta_\cC, 0)$.
Hereafter \eqref{eq:select_idx2} will be referred to as ``CS-PLIK". }

\section{Simulation Studies}
\label{sim}

The utility of the proposed methods was evaluated via extensive simulations. Denote by CS-MPLE,  a version of CoxCS that is based on the criteria of (\ref{eq:select_idx}). For completeness, we considered two other variations of CoxCS, namely,  CS-PLIK and  CS-Wald.
  The finite sample performance of  the proposed methods was compared with
the following  marginal screening methods designed for the survival data.

\begin{itemize}
\item CRIS: censored rank independence screening  proposed by~\cite{Song:2014}. 
\item  CORS: correlation screening, which is an extension of  sure independence screening to the censored  outcome data by using
inverse probability weighting; see~\cite{Song:2014}.
\item PSIS-Wald: Wald test  based on the  marginal Cox model fitted on  each covariate; see~\cite{DZhao:2012}.
\item PSIS-PLIK: partial likelihood ratio test based on the marginal Cox model fitted on each covariate, which is asymptotically equivalent to~\cite{DZhao:2012}.
\end{itemize}

%
%
%

 We illustrated our methods and compared them with the competing methods on data simulated as below.

\noindent \emph{Example}~1.
The survival time was generated from a Cox model with baseline hazards function being set to be  1, i.e.,
$$\lambda(t \mid \bfZ)=\exp(\bfbeta^{\rT}\bfZ),$$
where  $\bfZ$ were  generated from the standard normal distribution with equal correlation 0.5 and
$\bfbeta=(\bfone^{\rT}_5,-2.5,\bfzero^{\rT}_{p-6})^{\rT}$. 
\\
\\
\noindent \emph{Example}~2. 
The survival time  was generated from a Cox model with baseline hazards function being set to be  1, i.e.,
$$\lambda(t \mid \bfZ)=\exp(-1+\bfbeta^{\rT}\bfZ),$$ 
where  
$\bfbeta=(10,\bfzero_{p-2}^{\rT},1)^{\rT}$ and  all covariates were generated from the  independent standard normal distribution.

\noindent \emph{Example}~3.
The same as Example 2 except that the first $p-1$ covariates were generated from the multivariate standard normal distribution with an equal correlation of 0.9.


\begin{figure}[htbp] 
   \centering
   \includegraphics[width=6.0in]{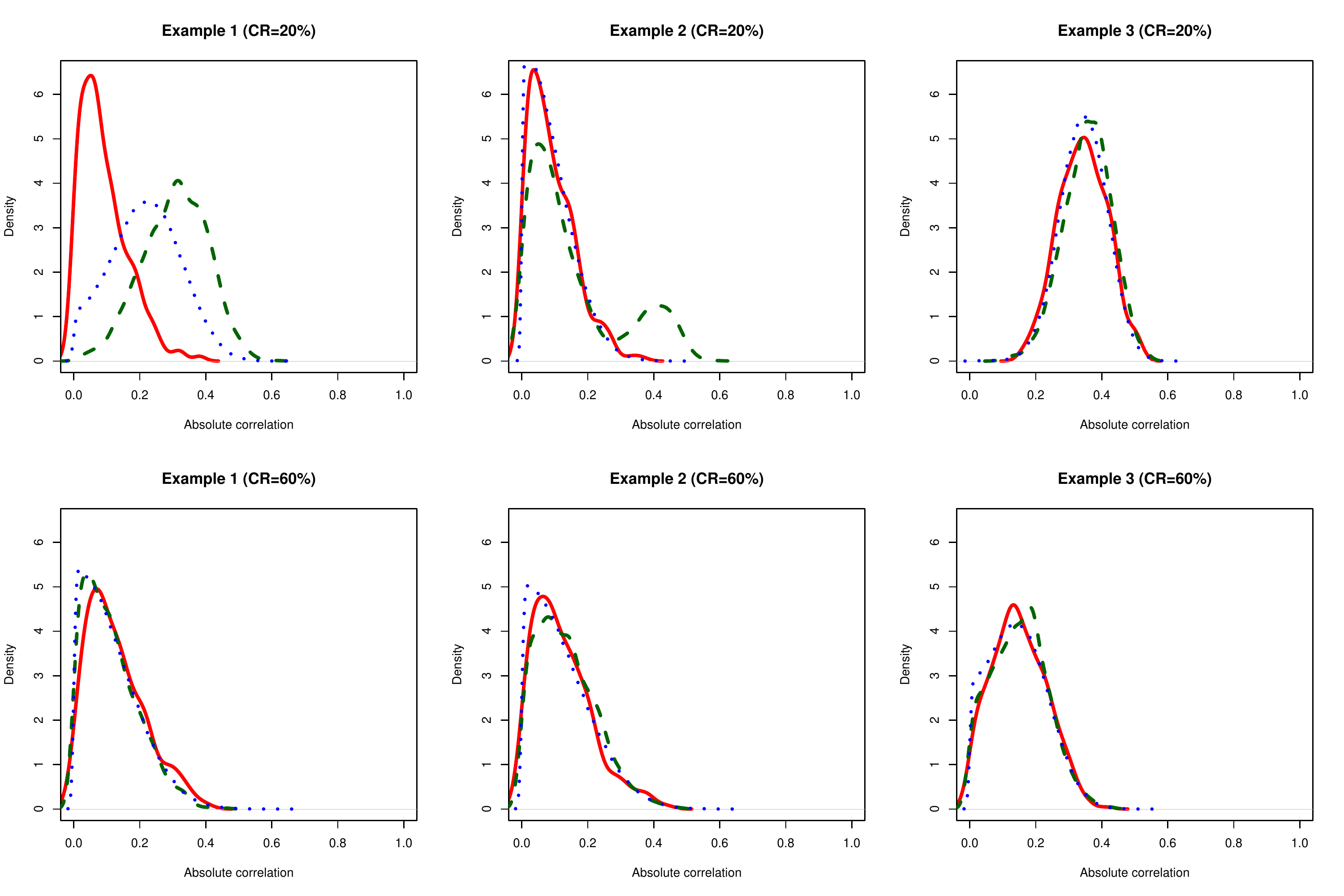} 
   \caption{\jian{Absolute correlation of the survival time and the covariate variables.} \grace{ The blue  short-dashed lines ($\cdot \cdot \cdot \cdot \cdot$) represents the distribution of the inactive variables;
the green long-dashed (-- -- --) lines for the active variables with relatively strong signals; the red solid lines (------) for the hidden active variable.}}
   \label{fig:abs_cor}
\end{figure}

The simulated data examples were designed in such a way that
variables $Z_6$ in Example 1 and $Z_p$ in Example in 3 possessed marginally weak but conditionally strong signals, which made   marginal screening approaches not
ideal for identifying them. For the GLM, \cite{Barut:2015} provided a similar simulation design in the context of non-censored regression. Figure~\ref{fig:abs_cor}  depicted  the distribution of the absolute correlation between  the survival time and the covariate variables,  where uncensored data were used to compute the marginal correlation between the event time and the covariate using an inverse probability weighting~\citep{Song:2014}. 
Clearly,  in Example 1 the marginal signal strength of $Z_6$ was  weaker than most noisy (inactive) variables ($Z_7-Z_{1000}$), while 
the marginal signal strength of $Z_{1000}$ in Examples 2 and 3 was similar or even lower than most noisy variables.
The marginal correlation between the survival time and each variable was getting weaker with heavier censoring.

In all these examples, the censoring times $C_i$ were independently generated from a uniform distribution $U[0,c]$, with $c$
chosen to  give approximately 20\% and 60\% of censoring proportions.   We set  $n=$ 100 and varied  $p$  from 1000 (high-dimensional) to 10000 (ultrahigh-dimensional). For each 
configuration, a total of 400 simulated datasets were generated.

\begin{figure}[htbp] 
   \centering
   \includegraphics[width=6.0in]{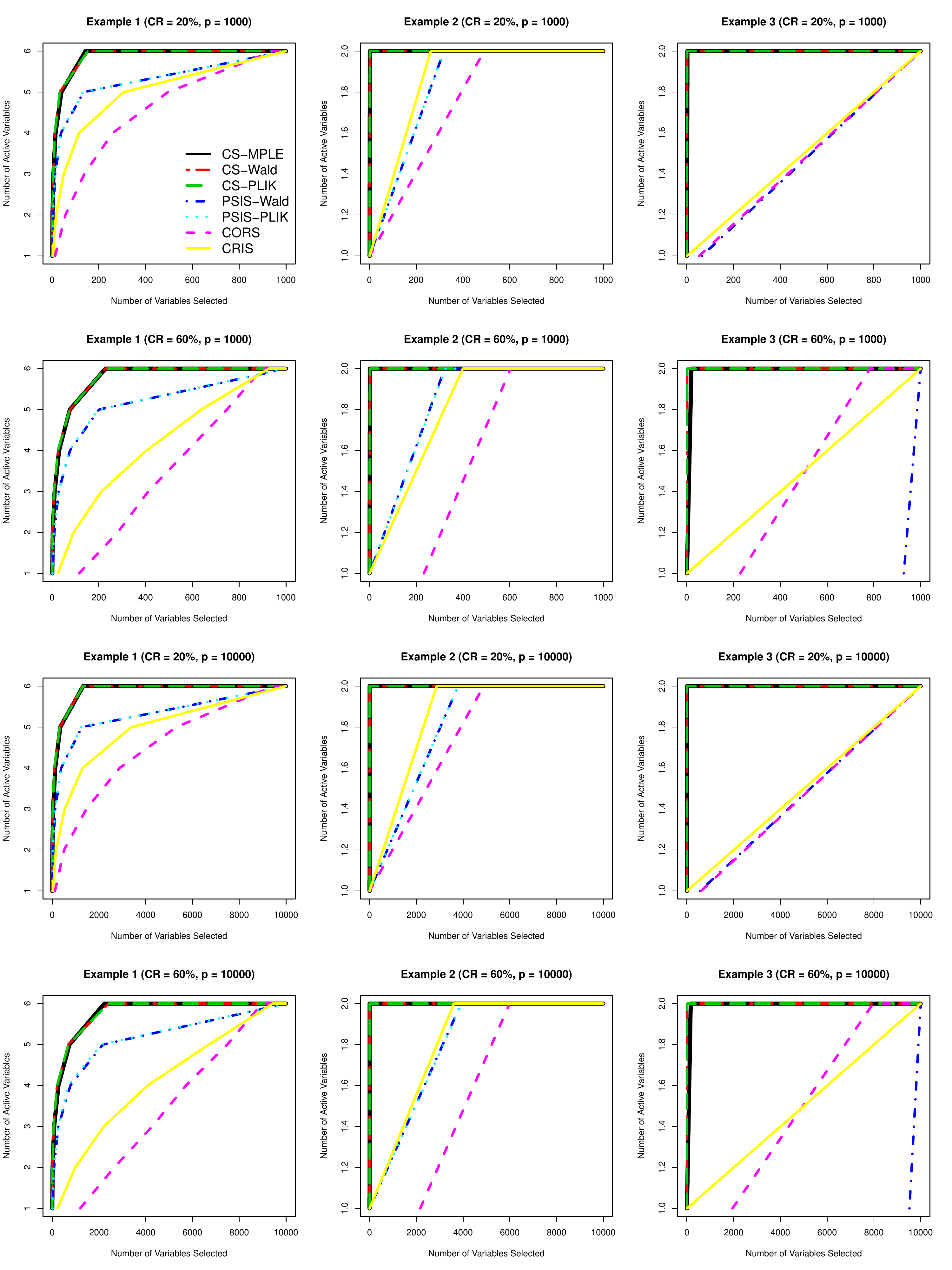} 
   \caption{\jian{Median number of active variables that are included in the model with different thresholds by different methods}}
   \label{fig:abs_cor}
\end{figure}

We considered two metrics to compare the performance between different methods:  the minimum model size (MMS) which is the  minimum number of variables that need to  be selected
in order to include all active variables, and the true positive rate (TPR) which is the proportion of active variables that are included in the first $n$ selected variables.   Hence, a method with small MMS and large TPR can be more efficient to  discover true signals. \jian{To have a fair comparison,} we added one (the number of conditioning variable in our examples) to MMS for the proposed methods (CS-MPLE, CS-Wald and CS-PLIK). In practice, identifying conditional sets normally requires some prior biological information. In our simulations, \jian{we simply choose the covariate $Z_1$ as the conditioning variable. In practice, } 
we propose to choose the variable with the  highest marginal signal strength as the conditioning variable, which can be a practical solution
in the absence of prior biological knowledge.  In Examples 1--3, $\cC=\{1\}$ is the true conditioning set, the signal of which was strong enough
to be easily selected by other marginal screening methods.

Table 1 demonstrated the superiority of our proposed methods under the difficult scenarios as reflected in Examples 1--3.  Indeed, the proposed methods drastically reduced MMS in Examples 1--3 compared to the marginal approaches.  Moreover,
we noted that all the  marginal screening methods had tremendous difficulties in identifying 
 $Z_6$ in Example 1 and $Z_{1000}$ in Examples 2--3.  Indeed, these variables had the lowest priorities to be included  by using the competing methods. 
On the other hand, the proposed approaches greatly outperformed the marginal approaches,  as the conditioning approaches effectively boosted the signal strengths of the ``hidden'' variables.  \jian{The performance by CS-MPLE, CS-Wald and CS-PLIK are quite similar in all the cases in Examples 1 and 2. In Example 3, there is a very high correlation among the covariate variables, CS-MPLE has a slightly larger MMS compared to the CS-Wald and CS-PLIK which well control the false discover rate in this cases. }  

\begin{table}[htbp]
\centering
\footnotesize
\begin{tabular}{rrr@{\hskip 2pt}lr@{\hskip 2pt}lr@{\hskip 2pt}lr@{\hskip 2pt}l}
  \toprule
  &&\multicolumn{4}{c}{$(n, p) = (100,1000)$}&\multicolumn{4}{c}{$(n, p) = (100, 10000)$}\\
 &Method & \multicolumn{2}{c}{MMS} &  \multicolumn{2}{c}{TPR} &  \multicolumn{2}{c}{MMS} &  \multicolumn{2}{c}{TPR} \\ 
  \midrule
  & CRIS & 1000.0 &(3.0) & 0.50 &(0.17) &9995.0 &(37.0) & 0.17 &(0.17) \\ 
& CORS & \ \ 944.0 &(168.2) & 0.33 &(0.33) &  9466.0 &(1101.8) & 0.00 &(0.17) \\ 
Example 1& PSIS-PLIK & 1000.0 &(0.0) & 0.67 &(0.17) & 10000.0 &(2.2) & 0.33 &(0.17) \\ 
CR $\approx$ 20\%& PSIS-Wald & 1000.0 &(0.0) & 0.67 &(0.17) & 10000.0\ &(2.2) & 0.33 &(0.17) \\ 
& CS-PLIK & 152.5 &(272.2) & 0.83 &(0.17) & 1322.0 &(2253.8) & 0.67 &(0.17) \\ 
& CS-Wald & 154.5 &(274.5) & 0.83 &(0.17) & 1286.5 &(2287.0) & 0.67 &(0.17) \\ 
& CS-MPLE & 143.0 &(249.0) & 0.83 &(0.17) & 1321.0 &(2305.8) & 0.50 &(0.17) \\[5pt]

& CRIS & 926.5 &(151.8) & 0.33 &(0.17) & 9429.0 &(1405.0) & 0.00 &(0.17) \\ 
 & CORS & 898.0 &(152.2) & 0.00 &(0.17) & 8976.0 &(1610.0) & 0.00 &(0.00) \\ 
 Example 1& PSIS-PLIK & 1000.0 &(2.0) & 0.67 &(0.17) & 9999.0 &(17.0) & 0.33 &(0.17) \\ 
 CR $\approx$ 60\%& PSIS-Wald & 1000.0 &(2.0) & 0.67 &(0.17) & 9999.0 &(17.0) & 0.33 &(0.17) \\ 
 & CS-PLIK & 227.0 &(351.2) & 0.83 &(0.17) & 2383.5 &(3331.5) & 0.50 &(0.33) \\ 
 & CS-Wald & 227.5 &(348.2) & 0.83 &(0.17) & 2403.5 &(3233.2) & 0.50 &(0.33) \\ 
 & CS-MPLE & 228.5 &(320.5) & 0.83 &(0.17) & 2229.5 &(3046.8) & 0.50 &(0.17) \\[5pt]
 
 & CRIS & 262.5 &(401.8) & 0.50 &(0.00) & 2871.0 &(4314.0) & 0.50 &(0.00) \\ 
 & CORS & 490.5 &(510.8) & 0.50 &(0.00) & 4878.5 &(5099.8) & 0.50 &(0.00) \\ 
Example 2 & PSIS-PLIK & 318.0 &(492.8) & 0.50 &(0.00) & 3777.0 &(5690.8) & 0.50 &(0.00) \\ 
CR $\approx$ 20\% & PSIS-Wald & 318.5 &(494.2) & 0.50 &(0.00) & 3786.5 &(5707.8) & 0.50 &(0.00) \\ 
 & CS-PLIK &2.0 &(0.0) & 1.00 &(0.00) & 2.0 &(0.0) & 1.00 &(0.00) \\ 
 & CS-Wald &2.0 &(0.0) & 1.00 &(0.00) & 2.0 &(0.0) & 1.00 &(0.00) \\ 
 & CS-MPLE & 2.0 &(0.0) & 1.00 &(0.00) & 2.0 &(0.0) & 1.00 &(0.00) \\[5pt]
  
 & CRIS & 399.5 &(478.0) & 0.50 &(0.00) & 3601.0 &(4894.2) & 0.50 &(0.00) \\ 
 & CORS & 603.5 &(390.5) & 0.00 &(0.00) & 5988.0 &(4353.0) & 0.00 &(0.00) \\ 
 Example 2& PSIS-PLIK & 325.0 &(498.5) & 0.50 &(0.00) & 3942.5 &(5679.5) & 0.50 &(0.00) \\ 
 CR $\approx$ 60\%& PSIS-Wald & 322.0 &(501.0) & 0.50 &(0.00) & 3915.5 &(5679.5) & 0.50 &(0.00) \\ 
 & CS-PLIK & 2.0 &(0.0) & 1.00 &(0.00) & 2.0 &(1.0) & 1.00 &(0.00) \\ 
 & CS-Wald & 2.0 &(0.0) & 1.00 &(0.00) & 2.0 &(2.0) & 1.00 &(0.00) \\ 
 & CS-MPLE & 2.0 &(0.0) & 1.00 &(0.00) & 2.0 &(4.0) & 1.00 &(0.00) \\[5pt]
 
 & CRIS & 1000.0 &(0.0) & 0.50 &(0.00) & 10000.0 &(0.0) & 0.50 &(0.00) \\ 
 & CORS & 1000.0 &(0.0) & 0.50 &(0.50) & 10000.0 &(0.0) & 0.00 &(0.00) \\ 
Example 3 & PSIS-PLIK & 1000.0 &(0.0) & 0.50 &(0.00) & 10000.0 &(0.0) & 0.50 &(0.00) \\ 
 CR $\approx$ 20\%& PSIS-Wald & 1000.0 &(0.0) & 0.50 &(0.50) & 10000.0 &(0.0) & 0.00 &(0.50) \\ 
 & CS-PLIK & 2.0 &(0.0) & 1.00 &(0.00) & 2.0 &(0.0) & 1.00 &(0.00) \\ 
 & CS-Wald & 2.0 &(0.0) & 1.00 &(0.00) & 2.0 &(0.0) & 1.00 &(0.00) \\ 
 & CS-MPLE & 3.0 &(4.0) & 1.00 &(0.00) & 7.5 &(33.0) & 1.00 &(0.00) \\[5pt]

 & CRIS & 1000.0 &(0.0) & 0.50 &(0.00) & 10000.0 &(0.0) & 0.50 &(0.00) \\ 
 & CORS & 783.0 &(463.8) & 0.00 &(0.50) & 7967.0 &(4972.2) & 0.00 &(0.00) \\ 
 Example 3 & PSIS-PLIK & 1000.0 &(0.0) & 0.50 &(0.00) & 10000.0 &(0.0) & 0.50 &(0.00) \\ 
 CR $\approx$ 60\%& PSIS-Wald & 1000.0 &(0.0) & 0.00 &(0.00) & 10000.0 &(0.0) & 0.00 &(0.00) \\ 
 & CS-PLIK & 2.0 &(0.0) & 1.00 &(0.00) & 2.0 &(0.0) & 1.00 &(0.00) \\ 
 & CS-Wald & 2.0 &(0.0) & 1.00 &(0.00) & 2.0 &(0.0) & 1.00 &(0.00) \\ 
 & CS-MPLE & 20.0 &(55.2) & 1.00 &(0.00) & 161.5 &(553.5) & 0.50 &(0.50) \\ 
   \bottomrule
\end{tabular}
\caption{\label{tab:1} \jian{Median minimum model size (MMS) and median true positive rates (TPR) along with their corresponding IQRs (in the parentheses) based on 400 simulated data sets.}}
\end{table}

\section{Application}
\label{app}

We illustrated the practical utility of the proposed method by applying it to analyze the diffuse large B-cell lymphoma (DLBCL)  dataset
of~\cite{Rosenwald:2002}. 
 The dataset, which was originally collected for identifying gene signatures relevant to the patient survival from time of chemotherapy,
included a total of 240 DLBCL patients with 138  deaths observed  during the followup and 
a median survival time of 2.8 years. Along with the clinical outcomes, the expression  levels of 7,399 genes were available for analysis. 
 In our subsequent analysis,  each gene expression  was  standardized to have mean zero and variance 1.

To facilitate the use of our method, we identified the conditional set  by resorting to the medical literature. 
As gene AA805575,  a
Germinal-center B-cell signature gene,  has been known to be predictive to DCBCL patients' survival in the literature \citep{Liu:2013, Gui:2005}, we used it as the
conditional variable in our proposed procedure. For comparisons, we also analyzed the same data using various competing methods introduced in the simulation section
and computed the corresponding concordance statistics (C-statistics)~\citep{Uno:2009}.  
  
Specifically, we randomly assigned 160 patients to the training set and 80 patients to  the testing set, while maintaining  the censoring proportion roughly the same in each set. 
For each split, we applied each method to select top 31($=160/\log(160))$ variables using the training set. LASSO was performed subsequently for refined modeling,  with the tuning parameter  selected by
the 10-fold cross-validation. The risk score for each subject was obtained by using the final model selected by LASSO in the training dataset  and the C-statistics
was obtained in the testing dataset.
A total of  100  splits were made and the average C-statistics and the model size (MS)  were reported in~Table \ref{tab:2}.  
By the criterion of  C-statistics, the proposed method seemed to have more predictive power.
\begin{table}[htbp]
\centering
\small
\begin{tabular}{lllllll}
\toprule
&CRIS& CORS& PSIS-PLIK&  PSIS-Wald& \\
\midrule
C-statistics&  0.54 (0.21) &  0.58 (0.20)    &  0.58 (0.19) &      0.55  (0.20) & \\
Model size&14.41 (3.00) & 6.83 (3.70)  & 15.22 (2.93)    &  15.65 (2.89)&  \\[2mm]
\midrule
       & CS-MPLE & CS-PLIK& CS-Wald \\\midrule
C-statistics &0.63 (0.18) & 0.63 (0.18)&   0.62 (0.19)\\
Model size & 16.74 (3.26) &15.90 (3.01) & 16.28 (3.41)\\
\bottomrule
\end{tabular}
\caption{Summary of C-statistics and the model size for different methods.}
\label{tab:2}
\end{table}

Our further scientific investigation focused on identifying the relevant genes by utilizing the full dataset. 
 Applying our proposed method, we selected top 44  ($=240/\log(240)$) genes, before using  LASSO to reach the final list.   It follows that CS-MPLE, CS-PLIK and CS-Wald selected 20,  16 and 16 genes, respectively.
Among the  22 uniquely selected genes by either of them, 14 genes were overlapped  and were reported in Table~\ref{tab:3}. \grace{
Twelve genes among these 22  genes belong to Lymph-node signature group, proliferation
signature group, and Germinal-center B-cell group defined by \cite{Rosenwald:2002}.
We observed  that 13 of these 22 genes  were  chosen by at least one of CRIS, CORS, PSIS-PLIK, and PSIS-Wald. On the other hand, gene
AB007866, Z50115, S78085, U00238, AL050283, J03040, U50196, and  AA830781, and M81695 were  only identified by using our methods.}

\grace {In fact, only a few studies  have suggested an important role of M81695~\citep{Deb:2003, Chow:2000, Mikovits:2001, Stewart:2000} or AA830781~\citep{Li2005BoostCox, Binder:2009, Schifano:2010} in predicting  DLBCL survival. Indeed,  as  the marginal correlation  between M81695 and the survival time  and between AA830781 and the survival time are markedly low at  0.008 and 0.097, respectively.  Thus,  it is highly likely to be missed by using the conventional  screening approaches.
  \cite{Schifano:2010}
 also commented  
their majorization-minimization algorithm selected  AA830781  because of coexpression or correlation with other relevant genes.  A more detailed investigation of its functions in the context of a broader class of blood cancers, including lymphoma,  may shed light on preventing, treating and controlling the lethal blood cancers. }

\begin{landscape}
\begin{table}[htbp]
\centering
\small
\begin{tabular}{lllccc}
\toprule
GenBank ID&Signature&Description&CS-MPLE & CS-PLIK& CS-Wald \\
\midrule
 LC{\_}25054& &&\checkmark& \checkmark& \checkmark\\
X77743&Proliferation&cyclin-dependent kinase 7 & \checkmark& \checkmark& \checkmark\\
U15552&&acidic 82 kDa protein mRNA& \checkmark& \checkmark& \checkmark\\
AB007866&&KIAA0406 gene product& \checkmark\\
BC012161& Proliferation& septin 1&\checkmark& \checkmark& \checkmark\\
 AF134159&  Proliferation&chromosome 14 open reading frame 1&\checkmark& \checkmark& \checkmark\\
 Z50115&Proliferation&thimet oligopeptidase 1&\checkmark\\
S78085&Proliferation&programmed cell death 2& \checkmark&\\
M29536&Proliferation &eukaryotic translation initiation factor 2&\checkmark& \checkmark& \checkmark\\
U00238& Proliferation&phosphoribosyl pyrophosphate &&\checkmark\\
&&amidotransferase\\
AL050283&Proliferation&sentrin/SUMO-specific protease 3& \checkmark\\
BF129543& Germinal-center-B-cell&ESTs, Weakly similar to A47224 &\checkmark& \checkmark& \checkmark\\
&&thyroxine-binding globulin precursor\\
M81695&&integrin, alpha X &\checkmark& \checkmark& \checkmark\\
D13666&Lymph node& osteoblast specific factor 2 (fasciclin I-like)&\checkmark& \checkmark& \checkmark\\
J03040&Lymph node&secreted protein& &\checkmark&\checkmark\\
U50196& &adenosine kinase&\checkmark& \checkmark& \checkmark\\
U28918& Proliferation&suppression of tumorigenicity 13 & \checkmark&& \checkmark\\
AA721746&&ESTs& \checkmark& \checkmark& \checkmark\\
AA830781& &&\checkmark& \checkmark& \checkmark\\
AF127481&&lymphoid blast crisis oncogene& \checkmark\\
M61906&&phosphoinositide-3-kinase&\checkmark& \checkmark& \checkmark\\
D42043&&KIAA0084 protein&\checkmark& \checkmark& \checkmark\\
\bottomrule
\end{tabular}
\caption{A comparison of genes that are selected by CS-MPLE, CS-PLIK and CS-Wald.}
\label{tab:3}
\end{table}
\end{landscape}

\section{Discussion}
\label{disc}
\jian{In this paper, we have proposed a new conditional variable screening approach for the Cox proportional hazard model  with ultra-high dimensional covariates. The proposed partial likelihood based conditional screening approaches are extremely computationally efficient, with a solid theoretical foundation. Our method and theory are extensions of the conditional sure independence screening (CSIS) \cite{Barut:2015}, which is designed for the GLM.  In the development of theory,  we introduce  the new concept of the conditional linear covariance for the first time, which is useful to specify the regularity conditions for  the model identifiability and the sure screening property. This also provides a solid building block  for a general theoretical framework of conditional variable screening in the context of other semi-parametric models, such as the partially linear single-index model.}

\grace{We have mainly focused on studying the theoretical properties of CS-MPLE, which extends the work of \cite{Barut:2015} in the GLM setting, though
development of the inference procedures  for the two variants of the proposed method, namely, CS-PLIK and CS-Wald, will be more involved and out of scope of this paper.  However,   as indicated by the simulation studies,   these two variants may induce substantial improvement especially when the variables are highly correlated. More research is warranted.  }

\jian{Our work also enlightens a few directions that are worth ensuing effort. First,  as our proposal requires the prior information to be known and informative,
it remains statistically challenging to develop efficient screening methods in the absence of such information.
 Recently, in the context of GLM, \cite{Hong:2015} has proposed a data-driven alternative when a pre-selected set of variables is unknown.
It is thus of substantial interest to develop a data-driven conditional screening for the survival model. Second, even with prior knowledge, an open question often lies in how to balance it with the information extracted from the given data. There has been some recent work 
 on how to incorporate  prior information. For example, 
 \cite{Wang:2013} 
 developed a LASSO method  by assigning different prior distributions to each subset according to a modified Bayesian information criterion that incorporates prior knowledge on both the network structure and the pathway information, and  \cite{Jiang:2015} proposed ``prior lasso" (plasso) to  balance between the prior information and the data. 
A natural extension of the current work is to develop a variable screening approach that incorporates more complex prior knowledge, 
such as the network structure or the spatial information of the covariates.  We will report the progress elsewhere.
}

\newpage
\bibliographystyle{spmpsci}
\bibliography{refs}

\section{Appendix}
\subsection{Proof of Theorem 1}
\begin{proof}
First we make the connection between $\beta_{j}$ to the expected conditional linear covariance between $Z_j$ and $\mP[\delta=1\mid \bfZ]$ given $\bfZ_\cC$, that is 
$$\mE[\mCov^*(Z_j, \mP[\delta=1\mid \bfZ]\mid \bfZ_{\cC})],$$ then by Condition \ref{cond:covariate}, we relate it to $\alpha_{j}$. For any $j\notin \cC$ and $k\in\cC$,   it is straightforward to see that
\begin{eqnarray}\label{eq:def_s}
s^{(m)}_{k}(t) = \mE[Z_{k}^{m}\lambda_0(t) \exp(\bfZ^{\rT}\bfalpha) S_T(t\mid\bfZ) S_C(t)],
\end{eqnarray}
and 
\begin{eqnarray}\label{eq:def_r}
r^{(m)}_{j,k}(t,\bfbeta_{\cC},\beta) =\mE[Z^m_{k}  \exp(\bfZ_{\cC}^{\rT}\bfbeta_{\cC} + Z_{j}\beta) S_T(t\mid\bfZ) S_C(t)],
\end{eqnarray}
for $m = 0,1$. Then
\begin{eqnarray}\label{eq:def_v_2}
\lefteqn{v_{j,k}(\bfbeta_{\cC},\beta) }\nonumber\\
&& = \int_0^\tau\mE\left[ W_{j,k}(t,\bfbeta_{\cC},\beta)\exp(\bfZ^{\rT}\bfalpha) S_T(t\mid\bfZ) S_C(t)\lambda_0(t)\right]\md t,
\end{eqnarray}
where 
\begin{eqnarray*}
W_{j,k}(t,\bfbeta_{\cC},\beta) = Z_k- \frac{\mE[Z_k \exp(\bfZ_{\cC}^{\rT}\bfbeta_{\cC} + Z_{j}\beta) S_T(t\mid\bfZ) S_C(t)]}{\mE[\exp(\bfZ_{\cC}^{\rT}\bfbeta_{\cC} + Z_{j}\beta) S_T(t\mid\bfZ) S_C(t)]}. 
\end{eqnarray*}
By Proposition \ref{prop:CLE}, 
\begin{eqnarray*}
\lefteqn{\mE\left[ W_{j,k}(t,\bfbeta_{\cC},\beta)\exp(\bfZ^{\rT}\bfalpha) S_T(t\mid\bfZ) S_C(t)\right]}\\
&&=\mE\left\{\mE^*\left[ W_{j,k}(t,\bfbeta_{\cC},\beta)\exp(\bfZ^{\rT}\bfalpha) S_T(t\mid\bfZ) S_C(t) \right]\right\}.
\end{eqnarray*}

 
By Definition \ref{def:cond_exp_v}, 
\begin{eqnarray*}
\lefteqn{v_j(\bfbeta_{\cC},\beta) = v_{j,j}(\bfbeta_\cC,\beta) -\sum_{k\in\cC} a_k v_{j,k}(\bfbeta_\cC,\beta)} \\
=&& \mE\left[ \mCov^*( Z_j, \mP[\delta=1\mid \bfZ] \mid \bfZ_{\cC})\right] - g(\bfbeta_{\cC},\beta). 
\end{eqnarray*}
where 
\begin{eqnarray*}
\lefteqn{ \mE\left[ \mCov^*( Z_j, \mP[\delta=1\mid \bfZ] \mid \bfZ_{\cC})\right]}\\
&& = \int_0^\tau\mE\left[(Z_j - \mE^*[Z_j \mid \bfZ_{\cC}])\exp(\bfZ^{\rT}\bfalpha) S_T(t\mid\bfZ) S_C(t)\lambda_0(t)\right]\md t, 
\end{eqnarray*}
and
\begin{eqnarray*}
\lefteqn{g_{j}(\bfbeta_{\cC},\beta)}\\
&&\qquad =  \int_0^\tau\frac{\mE[(Z_j - \mE^*[Z_j \mid \bfZ_{\cC}]) \exp(\bfZ_{\cC}^{\rT}\bfbeta_{\cC}+Z_j\beta) S_T(t\mid \bfZ)S_C(t)]}{\mE[\exp(\bfZ_{\cC}^{\rT}\bfbeta_{\cC}+Z_j\beta) S_T(t\mid \bfZ)S_C(t)]}\\
&&\qquad\qquad\qquad\qquad\qquad\qquad \times \mE\left[\exp(\bfZ^{\rT}\bfalpha) S_T(t\mid \bfZ)\lambda_0(t) S_C(t)\right]\md t. 
\end{eqnarray*}

By Definition \ref{def:pop_solution}, $\bfv_j(\bfbeta_{\cC,j},\beta_j) = \bfzero_{q+1}$, 
$$g_j(\bfbeta_{\cC,j}, \beta_j) = \mE\left[ \mCov^*( Z_j, \mP[\delta=1\mid \bfZ] \mid \bfZ_{\cC})\right].$$

When $\alpha_{j} = 0$, then $\mE\left[ \mCov^*( Z_j, \mP[\delta=1\mid \bfZ] \mid \bfZ_{\cC})\right] = 0$.
Thus $g_j(\bfbeta_{\cC,j},\beta_j) = 0$. 
Also, by Propositions \ref{prop:solution} and \ref{prop:CLE},  
$g_j(\bfbeta_{\cC,0},0) = 0$, then $\bfv_j(\bfbeta_{\cC,0},0) = \bfzero_{q+1}$. 
By uniqueness in Lemma~\ref{lem:unique}, $\beta_j = 0$.

When $\alpha_j \neq 0$,  by Condition \ref{cond:covariate}, we have 
$$
|g_j(\bfbeta_{\cC,j},\beta_j)| = | \mE\left[ \mCov^*( Z_j, \mP[\delta=1\mid \bfZ] \mid \bfZ_{\cC})\right]| > c_1 n^{-\kappa}.
$$
This implies that $g_j(\bfbeta_{\cC,j},\beta_j)$ and $ \mE\left[ \mCov^*( Z_j, \mP[\delta=1\mid \bfZ] \mid \bfZ_{\cC})\right]$ are both nonzero and have the same signs since they are equal.  Next we show for any $\bfbeta_{\cC}$, $g_j(\bfbeta_{\cC},0)$ and $\mE\left[ \mCov^*( Z_j, \mP[\delta=1\mid \bfZ] \mid \bfZ_{\cC})\right]$ have the opposite signs unless they are equal to zero. This fact implies that $\beta_j\neq 0$. Specifically, note that $\mP(\delta =  1\mid \bfZ)$ is the probability of occurring the event  and $S_T(t\mid \bfZ)S_C(t) = \mP(X > t \mid \bfZ)$ represents the probability at risk at time $t$. Based on Model~\eqref{eq:Cox_model}, for any $t$, 
$$\frac{\partial \mP(X > t \mid \bfZ)}{\partial Z_j} \times \frac{\partial \mP( \delta = 1 \mid \bfZ)}{\partial Z_j} \leq 0. $$
By Proposition~\ref{prop:cond_linear_cov}, 
$\mCov^*( Z_j,  \mP[\delta=1\mid \bfZ] \mid \bfZ_{\cC})$ and $\mCov^*[Z_j,  S_T(t\mid \bfZ)S_C(t)\mid \bfZ_\cC]$ have the opposite signs unless they are zero. This further implies that for any $\bfbeta_{\cC}$, 
\begin{eqnarray*}
\lefteqn{g_{j}(\bfbeta_{\cC},0) =  \int_0^\tau\frac{\mE[\exp(\bfZ_{\cC}^{\rT} \bfbeta_{\cC})\mCov^*[Z_j,  S_T(t\mid \bfZ)S_C(t)\mid \bfZ_\cC]]}{\mE[\exp(\bfZ_{\cC}^{\rT}\bfbeta_{\cC}) S_T(t\mid \bfZ)S_C(t)]}}\\
&&\qquad\qquad\qquad\qquad\qquad\qquad \times \mE\left[\exp(\bfZ^{\rT}\bfalpha) S_T(t\mid \bfZ)\lambda_0(t) S_C(t)\right]\md t.
\end{eqnarray*}
and $\mE\left[ \mCov^*( Z_j, \mP[\delta=1\mid \bfZ]\mid \bfZ_{\cC})\right]$ have opposite signs unless they are equal to zero. Therefore, $\beta_j \neq 0$.

\end{proof}

\subsection{Proof of Theorem 2}
\begin{proof}
For any $j\in \cM_{-\cC}$, we have $\beta_j \neq 0$ by Theorem \ref{thm:ident}, by mean value theorem, for some $\widetilde\beta_j \in (0, \beta_j)$, 
$$|v_{j}(\bfbeta_{\cC,j},0)| = |v_{j}(\bfbeta_{\cC,j},\beta_{j}) - v_{j}(\bfbeta_{\cC,j},0)| = \left|\frac{\partial v_{j}}{\partial \beta}(\bfbeta_{\cC,j},\widetilde\beta_j)\right||\beta_j|,$$
Next we show that $\left|\frac{\partial v_{j}}{\partial \beta}(\bfbeta_{\cC,j},\widetilde\beta_j)\right|$ is bounded.  For given any $\bfbeta_{\cC}$, consider $g_{j}(\bfbeta_{\cC},\beta)$ as a function of $\beta$, 
Then 
\begin{eqnarray*}
&&\frac{\partial g_j}{\partial \beta}(\bfbeta_\cC,\beta) =\mE\left[\int_0^{\tau}H_j(t,\bfbeta_{\cC},\beta)S_C(t)\md F_T(t\mid \bfZ)\right].
\end{eqnarray*}
where 
\begin{eqnarray*}
&&H_{j}(t, \bfbeta_\cC,\beta) = \frac{\mE[\exp(\bfZ_{\cC}^{\rT} \bfbeta_{\cC})\mCov^*[Z^2_j\exp(Z_j\beta),  S_T(t\mid \bfZ)\mid \bfZ_\cC]]}{\mE[\exp(\bfZ_{\cC}^{\rT}\bfbeta_{\cC}+Z_j\beta) S_T(t\mid \bfZ)]} \\
&&\qquad -\frac{\mE[\exp(\bfZ_{\cC}^{\rT} \bfbeta_{\cC})\mCov^*[Z_j\exp(Z_j\beta),  S_T(t\mid \bfZ)\mid \bfZ_\cC]]\mE[Z_j\exp(\bfZ_{\cC}^{\rT}\bfbeta_{\cC}+Z_j\beta) S_T(t\mid \bfZ)]}{[\mE[\exp(\bfZ_{\cC}^{\rT}\bfbeta_{\cC}+Z_j\beta) S_T(t\mid \bfZ)]]^2} \\
\end{eqnarray*}
By Condition \ref{cond:covariate}.1, $\mP(|Z|<K_0) = 1$, then $\sup_{\bfbeta_\cC,\beta}|H_j(t,\bfbeta_{\cC},\beta)|\leq 2 K_0^2$. Thus, 
$$\left|\frac{\partial v_{j}}{\partial \beta}(\bfbeta_{\cC,j},\widetilde\beta_j)\right|\leq \sup_{\bfbeta_\cC,\beta}\left|\frac{\partial g_j}{\partial \beta}(\bfbeta_\cC,\beta)\right| \leq 2K_0^2 |\mE[\mE[S_C(T)\mid \bfZ]] \leq 2K_0^2.$$
By the proof in Theorem \ref{thm:ident}, $g(\bfbeta_{\cC,j}, 0)$ and 
$\mE\left[ \mCov^*( Z_j, \mE\{F_{T}(C \mid \bfZ) \mid \bfZ\} \mid \bfZ_{\cC})\right]$ have the opposite signs, and by Condition \ref{cond:covariate}, 
$$|v_{j}(\bfbeta_{\cC,j},0)| 
 = | \mE\left[ \mCov^*( Z_j, \mP[\delta=1\mid \bfZ]\mid \bfZ_{\cC})\right]| + |g_{j}(\bfbeta_{\cC,j},0)| >  c_1 n^{-\kappa}.$$  
 Taking $c_2 = 0.5K_0^{-2} c_1$, $\beta_j > 0.5K_0^{-2} |v_j(\bfbeta_{\cC,j},0)| > c_2 n^{-\kappa} $. This completes the proof. 
\end{proof}
\subsection{Proof of Theorem 3}
\begin{proof}
For any $j\notin\cC$ and $k\in\cC\cup \{j\}$,  by \cite{lin1989robust}, we have 
$$\overline \bfV_j(\bfbeta_{\cC},\beta) = \mE_n\{\bfW_{i,j}(\bfbeta_{\cC},\beta)\} + o_p(1),$$ where $\mE_n[\cdot]$ denotes the empirical measure, which is defined as $\mE_n[\bfxi_i] =  n^{-1} \sum_{i=1}^n \bfxi_i$ for any random variables $\bfxi_1,\ldots, \bfxi_n$, and $\bfW_{i,j}(\bfbeta_{\cC},\beta)$ are independent over $i$, and write $\bfW_{i,j}(\bfbeta_{\cC},\beta) = [W_{i,j,k}(\bfbeta_{\cC},\beta),k\in\cC\cup\{j\}]^{\rT}$ with
\begin{eqnarray*}
\lefteqn{W_{i,j,k}(\bfbeta_{\cC},\beta) = \int_0^\tau \left\{Z_{i,k} - \frac{r^{(1)}_{j,k}(\bfbeta_{\cC},\beta, t)}{r^{(0)}_{j,k}(\bfbeta_{\cC},\beta,t)}\right\} \md N_i(t) }\\
&&\qquad \qquad- \int_0^\tau \frac{Y_i(t)\exp(\bfZ_{i,\cC}\bfbeta^{\rT}_{\cC}+ Z_{i,j}\beta)}{r^{(0)}_{j,k}(\bfbeta_{\cC},\beta,t)}\left\{Z_{i,k} - \frac{r^{(1)}_{j,k}(\bfbeta_{\cC},\beta, t)}{r^{(0)}_{j,k}(\bfbeta_{\cC},\beta,t)}\right\}\md \mE[N_i(t)]. 
\end{eqnarray*}
Note that given any $i, j, k$, with probability one $|W_{i,j,k}(\bfbeta_{\cC},\beta)|$ are uniformly bounded. Specifically, by Conditions \ref{cond:R}.2, \ref{cond:covariate}.1 and \ref{cond:beta_alpha},  with probability one, for all $t\in [0,\tau]$, $(\bfbeta_{\cC}^{\rT},\beta)^{\rT}\in \cB_j$,
$$ \left|Z_{i,k} - \frac{r^{(1)}_{j,k}(\bfbeta_{\cC},\beta, t)}{r^{(0)}_{j,k}(\bfbeta_{\cC},\beta,t)}\right| \leq |Z_{i,k}| + K_0,$$
 $$\left|\frac{Y_i(t)\exp(\bfZ_{i,\cC}\bfbeta^{\rT}_{\cC}+ Z_{i,j}\beta)}{r^{(0)}_{j,k}(\bfbeta_{\cC},\beta,t)}\right| \leq \exp\{K_0(K_1+\delta)-\log(L)\},$$
 and $$\left|\int_0^{\tau} \md \mE[N_i(t)]\right | \leq \Lambda_0(\tau)\exp(K_0K_1). $$
 Thus, with probability one, 
$$|W_{i,j,k}(\bfbeta_{\cC},\beta)| \leq K_2,$$
where $K_2 = 2K_0(1+\Lambda_0(\tau)\exp(2K_0K_1+K_0\delta-\log L))$. By the fact that  $\mE[W_{i,j,k}(\bfbeta_{\cC},\beta)] = 0$, 
$$\mVar[W_{i,j,k}(\bfbeta_{\cC},\beta)] = \mE[|W_{i,j,k}(\bfbeta_{\cC},\beta)|^2] < K_2^2$$

%
%
By Lemma 2.2.9 (Bernsterin's inequality) of~\cite{van1996weak}, for any $t>0$, for all $j, k$,  $\bfbeta_{{\cC}}$ and $\beta$, we have
\begin{eqnarray*}
\mP\left(|\mE_n(W_{i,j,k}(\bfbeta_{\cC},\beta))|>\frac{t}{n}\right) \leq 2 \exp\left(-\frac{1}{2}\frac{t^2}{n K^2_2+K_2 t/3}\right).
\end{eqnarray*}
Note that the above inequality holds for every $j\notin \cC$ and $k\in\cC\cup\{j\}$. By Bonferroni inequality, 
\begin{eqnarray*}
\mP\left(\|\mE_n(\bfW_{i,j}(\bfbeta_{\cC},\beta))\|_2>\frac{t}{(q+1)n}\right) \leq 2 (q+1)\exp\left(-\frac{1}{2}\frac{t^2}{n K^2_2+K_2 t/3}\right).
\end{eqnarray*}
Since, 
\begin{eqnarray*}
\|\overline\bfV_{j}(\bfbeta_{\cC},\beta)-\mE_n(\bfW_{i,j}(\bfbeta_{\cC},\beta))\|_2 = o_p(1).
\end{eqnarray*}
Then for any $\epsilon_1>0$ and $\epsilon_2>0$, there exits $N_1$, such that for any $n>N_1$
\begin{eqnarray*}
\mP(\|\overline\bfV_{j}(\bfbeta_{\cC},\beta)-\mE_n(\bfW_{i,j}(\bfbeta_{\cC},\beta))\|_2 > M\epsilon_1/2 ) < \epsilon_2.
\end{eqnarray*}
where $M$ is the same value in Condition~\ref{cond:beta_hat}. 
By Triangle inequality and Bonferroni inequality, we have
\begin{eqnarray*}
\lefteqn{\mP\left(\|\overline\bfV_{j}(\bfbeta_{\cC},\beta)\|_2 > \frac{t}{(q+1)n} \right)}\\
&& \leq \mP\left(\|\mE_n(\bfW_{i,j}(\bfbeta_{\cC},\beta))\|_2 > \frac{t}{(q+1)n} -M\epsilon_1/2\right) + \mP(\|\overline\bfV_{j}(\bfbeta_{\cC},\beta)-\mE_n(\bfW_{i,j}(\bfbeta_{\cC},\beta))\|_2 > M\epsilon_2/2 ) 
\end{eqnarray*}
When $n\to\infty$, take $t = c_2M(q+1)n^{1-\kappa}/2>0$ on both side of the inequality, where $c_2$ is the same value in Theorem \ref{thm:beta_low_bound}, we have 
\begin{eqnarray*}
\mP\left(\|\overline\bfV_{j}(\bfbeta_{\cC},\beta)\|_2 > \frac{Mc_2}{2}(n^{-\kappa} -\epsilon_1)\right) \leq 2 (q+1)\exp\left(-\frac{M^2c_2^2}{8(q+1)^2}\frac{n^{1-2\kappa}}{K^2_2+K_2 n^{-\kappa}/3}\right)+\epsilon_2.
\end{eqnarray*}
Take $N =\max\{ \lceil (K_2/3)^{1/\kappa}\rceil,N_1\}$, then for any $n>N$, $n^{-\kappa}<3/K_2$, and 
\begin{eqnarray*}
\mP\left(\|\overline\bfV_{j}(\bfbeta_{\cC},\beta)\|_2 > \frac{Mc_2}{2}(n^{-\kappa} -\epsilon_1)\right) \leq 2 (q+1)\exp\left(-\frac{M^2c_2^2}{8(q+1)^2}\frac{n^{1-2\kappa}}{K^2_2+1}\right)+\epsilon_2.
\end{eqnarray*}

Note that the above inequality holds for all $(\bfbeta_{\cC}^{\rT},\beta)^{\rT} \in \cB_j$,  particularly for $(\bfbeta_{\cC,j}^{\rT},\beta_j)^{\rT}$, $j \notin \cC$. Also, we have $\overline\bfV_{j}(\widehat\bfbeta_{\cC,j},\widehat\beta_j) = \bfzero_{q+1}$. By Condition~\ref{cond:beta_hat}, we have
\begin{eqnarray*}
\lefteqn{\mP\left(|\widehat\beta_j-\beta_j| >\frac{c_2}{2}(n^{-\kappa} -\epsilon_1)\right) }\\
&&\leq \mP\left(\|(\widehat\bfbeta_{\cC,j}^{\rT},\widehat\beta_j)^{\rT} -(\bfbeta_{\cC,j}^{\rT},\beta_j)^{\rT} \|_2 >\frac{c_2}{2}(n^{-\kappa} -\epsilon_1)\right) \\
&&\leq 2 (q+1)\exp\left(-\frac{M^2c_2^2}{8(q+1)^2}\frac{n^{1-2\kappa}}{K^2_2+1}\right)+\epsilon_2.
\end{eqnarray*}
Taking $c_3 = \frac{M^2c_2^2}{8(q+1)^2(K^2_2+1)}$ and by Bonferroni  completes the proof for part 1.

For part 2, by Theorem~\ref{thm:beta_low_bound}, 
$$\min_{j\in\cM_{-\cC}}|\beta_j| > c_2 n^{-\kappa}.$$
Note that, for any $j\in\cM_{-\cC}$, event
\begin{eqnarray*}
\lefteqn{\left\{|\widehat\beta_j - \beta_j| \leq c_2 n^{-\kappa}/2 - \epsilon_1\right\}}\\
&&\subseteq \left\{|\widehat\beta_j| \geq  |\beta_j| - c_2 n^{-\kappa}/2+\epsilon_1\right\}\\
&&\subseteq \left\{|\widehat\beta_j| \geq  c_2 n^{-\kappa}/2+\epsilon_1\right\}. 
\end{eqnarray*}
Take $\gamma_n = c_4 n^{-\kappa}$ with $c_4 = c_2/4$, 
\begin{eqnarray*}
\lefteqn{\left\{\max_{j\in\cM_{-\cC}}|\widehat\beta_j - \beta_j| \leq c_2 n^{-\kappa}/2 - \epsilon_1\right\}}\\
&&\subseteq\left\{\min_{j\in\cM_{-\cC}}|\widehat\beta_j|   \geq c_2 n^{-\kappa}/2+\epsilon_1\right\} \\
&&\subseteq\left\{\min_{j\in\cM_{-\cC}}|\widehat\beta_j|   \geq \gamma_n +\epsilon_1\right\}.  
\end{eqnarray*}
Thus, 
\begin{eqnarray*}
\lefteqn{\mP\left[\cM_{-\cC} \subseteq \widehat\cM_{-\cC}\right]}\\
&&= \mP\left[\min_{j\in \cM_{-\cC}}|\widehat \beta_j| > \gamma_n \right]\\
&&\geq \mP\left[\min_{j\in \cM_{-\cC}}|\widehat \beta_j| > \gamma_n +\epsilon_1\right] \\
&&\geq 1 - \mP\left[\max_{j\in\cM_{-\cC}}|\widehat\beta_j - \beta_j| \leq c_2 n^{-\kappa}/2 - \epsilon_1\right]\\
&&\geq 1 -2w(q+1)\exp(-c_3 n^{1-2\kappa}) - \epsilon_2. 
\end{eqnarray*}
Let $n\to \infty$, we have for any $\epsilon_2>0$, 
\begin{eqnarray*}
\lim_{n\to\infty}\mP\left[\cM_{-\cC} \subseteq \widehat\cM_{-\cC}\right] \geq 1 - \epsilon_2.
\end{eqnarray*}
Note that the left side of the above equation does not depends on $n$ any more. Taking $\epsilon_2\to 0$ completes proof.

%


%

\end{proof}

\end{document}